\documentclass[twocolumn,showpacs,preprintnumbers,amsmath,amssymb,prl,superscriptaddress]{revtex4-1}
\usepackage{graphicx}
\usepackage{amsmath}
\usepackage{amssymb}
\usepackage{dcolumn}
\usepackage{color}
\usepackage{subeqnarray}
\usepackage[utf8]{inputenc}
\usepackage[english]{babel}
\usepackage{mathrsfs}
\usepackage{wasysym}

\begin{document}

\preprint{}

\title{ Anti-Parity-Time Symmetric Optical Four-Wave Mixing in Cold Atoms}

\author{ Yue Jiang}\thanks{These authors contributed equally to this work.}
\affiliation{Department of Physics, The Hong Kong University of Science and Technology, Clear Water Bay, Kowloon, Hong Kong, China}

\author{Yefeng Mei}\thanks{These authors contributed equally to this work.}
\affiliation{Department of Physics, The Hong Kong University of Science and Technology, Clear Water Bay, Kowloon, Hong Kong, China}

\author{Ying Zuo}
\affiliation{Department of Physics, The Hong Kong University of Science and Technology, Clear Water Bay, Kowloon, Hong Kong, China}

\author{Yanhua Zhai}
\affiliation{Department of Physics, Kennesaw State University, Marietta, Georgia 30060, USA}

\author{Jensen Li}
\affiliation{Department of Physics, The Hong Kong University of Science and Technology, Clear Water Bay, Kowloon, Hong Kong, China}

\author{Jianming Wen}\email{Corresponding author: jianming.wen@kennesaw.edu}
\affiliation{Department of Physics, Kennesaw State University, Marietta, Georgia 30060, USA}

\author{Shengwang Du}\email{Corresponding author: dusw@ust.hk}
\affiliation{Department of Physics, The Hong Kong University of Science and Technology, Clear Water Bay, Kowloon, Hong Kong, China}

\date{\today }

\begin{abstract}
Non-Hermitian optical systems with parity-time (PT) symmetry have recently revealed many intriguing prospects that outperform conservative structures. The prevous works are mostly rooted in complex arrangements with controlled gain-loss interplay. Here, we demonstrate anti-PT symmetry inherent in nonlinear optical interactions based upon forward optical four-wave mixing in a laser-cooled atomic ensemble with negligible linear gain and loss. We observe the pair of frequency modes undergo a nontrivial anti-PT phase transition between coherent power oscillation and optical parametric amplification in presence of a large phase mismatch. 
\end{abstract}

\maketitle

Symmetries, as the fundamental properties of nature, play an essential role in our understanding of the universe. Recently, the notion of parity-time (PT) symmetry \cite{2,3,4,5,6} has started to attract significant attention owing to its potential for novel optical effects that are unattainable with usual Hermitian systems. For instance, coupled optical waveguides with balanced gain and loss can be used to steer optical beam profiles inaccessible with conservative architectures\cite{7}. The origin of this stems from the counterintuitive observation by Bender and Boettcher \cite{8} that certain non-Hermitian Hamiltonians, commuting with the joint parity and time-reversal operator, can possess entirely real spectra below some phase-transition point or exceptional point (EP). At EP \cite{8,9,10,11}, PT symmetry spontaneously breaks down and the system undertakes a new phase that admits conjugate pairs of complex eigenvalues. Despite much theoretical success of developing PT symmetry in quantum theory \cite{12,13}, yet, the search of such a quantum Hamiltonian remains highly elusive in the real physical world.

Owing to the mathematical equivalence between quantum Schrödinger equation and paraxial optical wave propagation equation, PT symmetry was first brought into the realm of optics by noticing that non-Hermitian PT-symmetric potentials can be easily realized through spatially modulating complex refraction \cite{14}. Indeed, subsequent works have weakened their relevance of quantum origin by typically observing striking PT phase transitions in various optical settings via interleaving balanced gain and loss regions. As shown by many works, such segmented arrangements can result in a variety of interesting phenomena \cite{15,16,17,18,19,21,22,25,26} that do not necessarily have corresponding counterparts in conservative systems.

As a counterpart, anti-PT symmetry \cite{27,28,30,31,32,35} has recently aroused great interest in the community. Contrary to PT symmetry, an anti-PT-symmetric Hamiltonian anti-commutes with the combined parity-time operator \cite{27}. This anti-commutation relationship suggests that properties of anti-PT systems would form conjugate counterparts to those of PT systems, yet allowing the appearance of EPs. Complementary to PT symmetric systems, indeed, noteworthy effects have been revealed including constant refraction \cite{27,30}, coherent switch \cite{31}, asymmetric-mode switch \cite{32}, and energy-difference conserving dynamics \cite{29}. In contrast to a PT system, of importance, an anti-PT system does not rely on the interplay of gain and loss to exploit non-Hermitian dynamics. Consequently, this characteristics may play a crucial role in realizing non-Hermitian dynamics towards the quantum domain without Langevin noises \cite{50}.

In spite of these impressive progress, most studies insofar have been centered on synthetic optical structures with spatially separated loss and/or gain distributions in order to create PT/anti-PT-symmetric potentials in guiding or scattering input light. As a result, this confines nearly all the demonstrations to either linear optics \cite{2,3,4,5,11,15} or single-mode nonlinear dynamics \cite{6,37}. Moreover, fabricating compound photonic structures turns out to be technically challenging in part to physical limitation of gain materials plus the codependency between the real and imaginary parts of refractive-index landscapes ruled by Kramers-Kronig relations. These concerns have led us to ask whether PT/anti-PT symmetry exists in any natural setting.

In this Letter, we demonstrate that nonlinear optical wave mixings \cite{1} can serve as a fertile arena for the exploration of innate PT/anti-PT symmetry between two distinct parametric signals. In particular, we find that anti-PT symmetry arises naturally in forward four-wave mixing (FFWM) for the pair of parametric fields, Stokes and anti-Stokes. Specifically, in FFWM without linear Raman gain and loss the evolution of paired Stokes and anti-Stokes modes exhibits perfect anti-PT in presence of phase mismatch. We here report the first observation of anti-PT-symmetric FFWM with negligible linear gain and loss in laser-cooled atoms. 

\begin{figure} [t]
\centering
\includegraphics[width=8.5 cm]{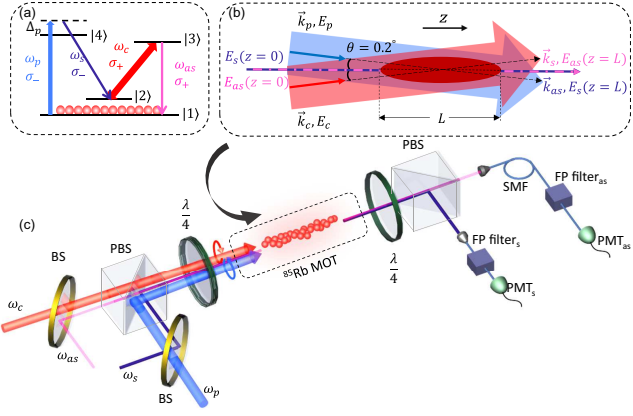}
\caption{(color online) Anti-PT symmetric forward four-wave mixing (FFWM) in a two-dimensional (2D) magneto-optical trap (MOT) of $^{85}$Rb atoms. (a) $^{85}$Rb atomic energy-level diagram: $|1\rangle =|5S_{1/2}, F=2\rangle $, $|2\rangle =|5S_{1/2}, F=3\rangle $, $|3\rangle =|5P_{1/2}, F=3\rangle $ and $|4\rangle =|5P_{3/2}, F=3\rangle $. (b) Geometrical arrangement on four interacting fields inside the MOT, where both pump ($E_p$) and coupling ($E_c$) beams of the same diameter of 1.54 mm are twisted by an angle of $\theta \mathrm{=0.}{\mathrm{2}}^{{}^\circ }$ to $\mathrm{+}z$ axis. The Stokes ($E_s$) and anti-Stokes ($E_{as}$) light collinearly propagate along the $\mathrm{+}z$ axis with the same waist diameter of 300 $\mathrm{\mu}$m at the MOT center. (c) Schematic of the experimental setup. In the injection part, the coupling (${\omega }_c$) and anti-Stokes (${\omega }_{as}$) beams, same to the pump (${\omega }_p$) and Stokes (${\omega }_s$) beams, are first combined by a beam splitter (BS) and polarization beam splitter (PBS) and then directed through a quarter wave plate (QWP) to generate the desired circular polarizations. In the collection part, the output Stokes and anti-Stokes beams are first separated through their polarizations and then measured by photomultiplier tubes (${\mathrm{PMT}}_s$ and ${\mathrm{PMT}}_{as}$) after the narrowband Fabre-Perot (FP) filters.}
\label{fig:figure1}
\end{figure}

\begin{figure} [t]
	\includegraphics[width=7 cm]{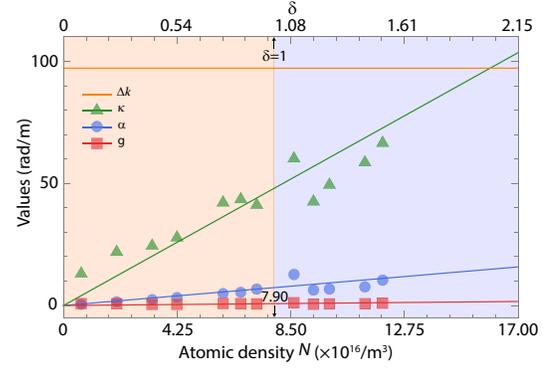}
	\caption{(color online) Comparison of phase mismatch $\mathrm{\Delta }k$, nonlinear coupling coefficient $\kappa$, linear Raman gain coefficient $g$ of the Stokes field, and linear absorption coefficient $\alpha $ of the anti-Stokes field as functions of atomic density $N$ or $\mathrm{\delta }$ at the two-photon resonance. The shaded color areas mark two different phases of anti-PT symmetry.}
\label{fig:fig2}
\end{figure}

As schematically shown in Fig. \ref{fig:figure1}(a), FFWM \cite{HarrisPRL1996, 44} occurs in a cold atomic ensemble with double-$\Lambda$ four-level configuration. A pump laser (\(\omega _{p} \)) is blue detuned by  \(\Delta _{p}\)  from the atomic transition   $|1\rangle \rightarrow |4 \rangle$ and a weak Stokes field  ($\omega _{s}$) follows  $|4\rangle\rightarrow  |2 \rangle$. Another coupling laser (\(\omega _{c}\)) is on resonance at  $|2 \rangle  \rightarrow  |3 \rangle$  and renders the atomic medium transparent \cite{45,46} for the weak anti-Stokes field (\(\omega _{as}\)), which is on the transition $|3 \rangle  \rightarrow  |1 \rangle$. We adopt a large  $\Delta _{p}=2\pi \times 90$ MHz  so that the atomic population primarily resides in the lower ground state  $|1\rangle$. Under this approximation, the pump and coupling beams are nearly undepleted. For a continuous-wave FWM process, energy conservation requires  \(\omega _{p}+ \omega _{c}= \omega _{s}+ \omega _{as}\), or  \(\omega _{p}- \omega _{s}= \omega _{as}- \omega _{c}\). We hence work under the two-photon resonance condition  \(  \Delta  \omega = \omega _{p}- \omega _{s}- \omega _{21}= \omega _{as}- \omega _{c}- \omega _{21}=0 \), where  \(  \omega _{21} \)  is the frequency difference between two ground states  $|1 \rangle  \rightarrow  |2 \rangle$. We next carefully select the pump power to ensure the linear Raman gain negligible to the Stokes field. For zero dephasing rate (\(  \gamma _{12}=0 \)) between  $|1 \rangle$  and  $|2 \rangle$, perfect electromagnetically induced transparency (EIT)\cite{45,46} caused by the coupling light allows anti-Stokes lossless propagation. In order to observe the anti-PT phase transition, we employ the geometric arrangement for four interacting beams as illustrated in Fig. \ref{fig:figure1}(b), where the twisted pump and coupling beams intersect with the collinear anti-Stokes and Stokes fields at a small angle  \(  \theta \) to the principal  \( +z \)  axis. The experimental setup is depicted in Fig. 1(c). Under these considerations, one can show that the slowly-evolved conjugate Stokes and anti-Stokes field amplitudes, \(  E_{s}^{\ast} \) and \( E_{as} \), obey Schrödinger-like coupled equations
\begin{eqnarray}
i\frac{\partial}{\partial z} \left[\begin{array}{c} E_{as} \\ E_s^* \end{array} \right] = \mathcal{H} \left[ \begin{array}{c} E_{as} \\ E_s^* \end{array} \right]
\label{eq:SchrodingerEq1}
\end{eqnarray}
where the effective Hamiltonian
\begin{eqnarray}
\mathcal{H}=\begin{bmatrix} -\Delta k/2 & -\kappa \\ \kappa & \Delta k/2 \end{bmatrix}.
\label{eq:H}
\end{eqnarray}
\(  \Delta k=k_{as}+k_{s}- \left( k_{p}+k_{c} \right) cos \theta  \) is the real phase mismatch with four involved wavenumbers  \( k_{j}  \left( j=s,as,p,c \right)  \)  in vacuum, and  \(  \kappa  \)  denotes the real nonlinear coupling coefficient. From Eq. (\ref{eq:SchrodingerEq1}), we obtain a pair of eigen-propagation constants
\begin{equation}
\lambda_{\pm}=\pm\lambda=\pm\frac{\Delta k}{2}\sqrt{1-\delta^2},
\label{eq:eigenvalueEq2}
\end{equation}
where  \(  \delta = \vert \frac{ \kappa }{ \Delta k/2} \vert  \). Here, the real eigenvalues, resembling the phase constant, define the rate at which the phase changes as the wave propagates; whereas the imaginary ones, referring to as the attenuation (or amplification) constant. Equations (\ref{eq:SchrodingerEq1}) and (\ref{eq:eigenvalueEq2}) contain interesting physics. First of all, for this  \( 2 \times 2 \)  matrix,  \( \mathcal{H} \)  anti-commutes with the combined parity-time ($\mathcal{P}\mathcal{T}$) operator,  $\left\{\mathcal{H} ,\mathcal{P}\mathcal{T}\right\} =0$, signifying a new format of anti-PT Hamiltonian. Here  \( P \)  means switching seeding operation (\( P \) :  \( E_{as} \rightarrow E_{s}^{*} \) ,  \( E_{s}^{*} \rightarrow E_{as} \)) while  \( T \)  complex conjugation (\( T \) :  \( E_{as} \left( z \right)  \rightarrow E_{as}^{*} \left( -z \right)  \) ,  \( E_{s}^{*} \left( z \right)  \rightarrow E_{s} \left( -z \right)  \)). Although  \( \mathcal{H}  \)  looks different from previous ones \cite{27,29,30,31,32,33,34,35}, they share some common features. Except for the similar diagonal terms, both off-diagonal ones belong to anti-Hermitian coupling and their product is thereby accompanied by a \textit{negative} sign. Unlike the beam-splitter-type interaction \cite{4,27,29,30,31,32,34,35}, the twin-field evolution here follows the Bogoliubov transformation \cite{33,43}. The solution of Eq.~(\ref{eq:SchrodingerEq1})  at  \( z=L \)  can be accordingly expressed as linear superposition of exponential functions of the form  \( e^{ \pm i \lambda z} \)  in terms of the inputs at  \( z=0 \) , which is
\begin{eqnarray}
\left[\begin{array}{c} E_{as}\left(z\right) \\ E_s^*\left(z\right) \end{array} \right] = \begin{bmatrix} A & B \\ C & D \end{bmatrix} \left[ \begin{array}{c} E_{as}\left(0\right) \\ E_s^*\left(0\right) \end{array} \right]
\label{eq:TransferMatrixEq3}
\end{eqnarray}
with \( A, B, C\) and \( D \)  being the input-output transmission functions. Specifically, in the symmetry-breaking regime (\(  \delta <1 \)) with real  \(  \pm  \lambda  \), one has  \( A=D^{\ast}=cos \left(  \lambda L \right) +i\frac{ \Delta k}{2 \lambda }sin \left(  \lambda L \right)  \) and \( B=C^{\ast}=i\frac{ \kappa }{ \lambda }sin \left(  \lambda L \right)  \). In this region, the Stokes and anti-Stokes output powers are bounded and oscillate coherently. Contrarily, when  \(  \delta >1 \), the interaction is governed by anti-PT symmetry and the two eigenvalues become a complex conjugate pair (\(  \pm  \lambda = \pm i \gamma  \)). In other words, one eigenstate undertakes gain while the other vanishes after some propagation distance. Accordingly, the transmission functions transform to  \( A=D^{*}=cosh \left(  \gamma L \right) +i\frac{ \Delta k}{2 \gamma }sinh \left(  \gamma L \right)  \)  and  \( B=C^{*}=i\frac{ \kappa }{ \gamma }sinh \left(  \gamma L \right)  \). For  \( L\rightarrow\infty \), because of the disappearance of one eigenmode, we have  \(  \vert A \vert ^{2}= \vert C \vert ^{2}= \vert B \vert ^{2}= \vert D \vert ^{2} \), implying the independence of the remaining output mode on the input. Alternatively, in the phase-unbroken case, the system behaves as optical parametric amplifier, a result of coherent power transfer from input pump and coupling to paired Stokes and anti-Stokes. The phase transition occurs at the EP ( \(  \delta =1 \)) where both eigenvalues and eigenvectors coalescence together. Aside from these compelling properties, we remark again that unlike previous research focus on the propagation of input light traversing non-Hermitian systems, here the creation of paired modes is intrinsically companioned with anti-PT symmetry which has not been noticed before.

\begin{figure} [t]
	\includegraphics[width=7 cm]{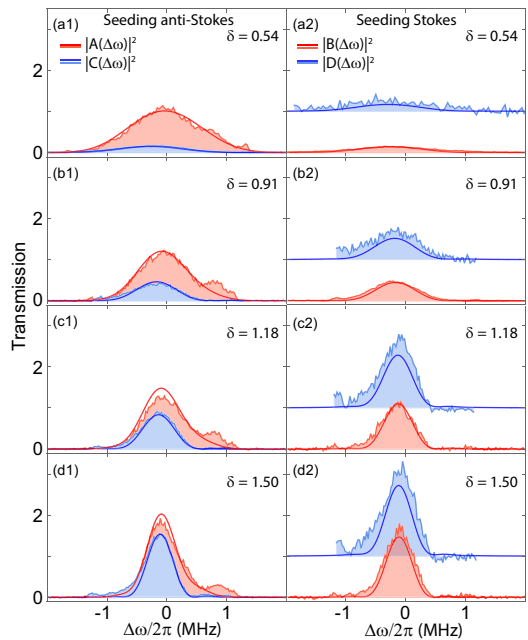}
	\caption{(color online) Transmission spectra of Stokes (blue) and anti-Stokes (orange) outputs at different $\delta$ for seeding only anti-Stokes, (a),  and only Stokes, (b). The theoretical simulations (solid lines) agree well with the recorded data (shaded). The experimental parameters here are $\Omega_p=2\pi\times 5.0$ MHz, $\Omega_c=2\pi\times 8.7$ MHz, $\Delta_p=2\pi\times 8.7$ MHz, $\gamma_{12}=2\pi\times 0.015$ MHz, and $\Delta k=97.2$ rad/m.}
\label{fig:fig3}
\end{figure}
\begin{figure} [t]
	\includegraphics[width=0.48\textwidth]{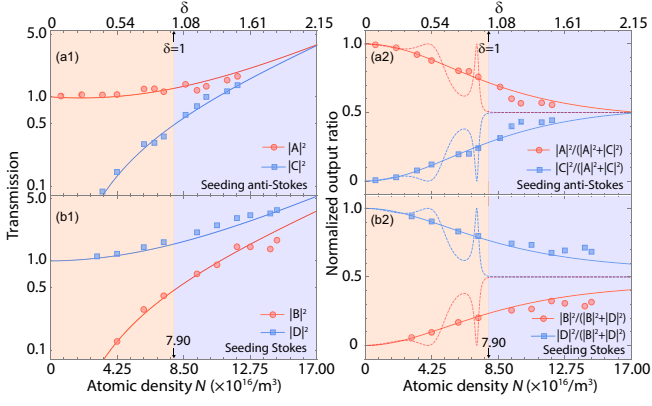}
	\caption{(color online) Evolution of anti-PT supermodes in terms of normalized output powers as functions of atomic density \(  N \)  or  \(  \delta  \)  at two-photon resonance (\(  \Delta  \omega =0 \)) under different seeding. The solid lines are theoretical curves, and circles and squares are the experimental data. The shaded areas are anti-PT phase broken (light orange) and phase unbroken (light purple) regimes. The dashed lines in (a2) and (b2) are numerical simulations for a medium with a longer length ($10L$) so as to explicitly reveal the abrupt phase transition. Other involved parameters are same as those in Fig.~\ref{fig:fig3}.}
\label{fig:fig4}
\end{figure}
\begin{figure} [t]
	\includegraphics[width=0.48\textwidth]{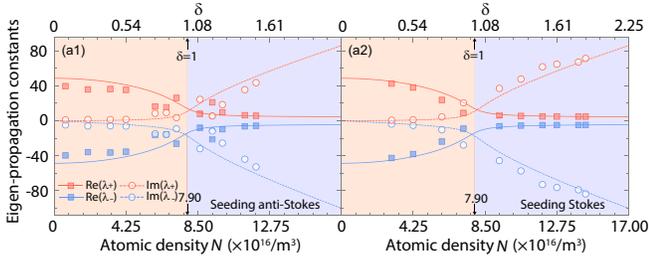}
	\caption{(color online) Re $\left[  \lambda _{ \pm } \right]$  and Im$\left[  \lambda _{ \pm } \right]$ of the two eigen-propagation constants of paired Stokes-anti-Stokes supermodes as a function of trapped atomic density \(  N \)  or  \(  \delta  \)  subject to different seeding operations.}
\label{fig:fig5}
\end{figure}

To confirm our theory, we work with laser-cooled $^{85}$Rb atoms in a dark-line two-dimensional (2D) magneto-optical trap (MOT) \cite{40} with $L=1.5$ cm [Fig.~\ref{fig:figure1}(c)]. We run the experiment periodically at a fixed FWM duty window (500 $ \mu $s) in each cycle but a varying repetition rate (in the range of 200 Hz$ \sim $1700 Hz). The repetition rate change conversely modifies the available time for the MOT loading, which thereupon alters the trapped atomic density  \( N \)  so as  \(  \kappa  \). By this way, with the fixed  \( L \), we are capable of tuning  \( N \)  from  \( 0.80 \times 10^{16} \) m$^{-3}$ to  \( 14.50 \times 10^{16} \)  m$^{-3}$. The independence of  \(  \Delta k \)  on  \( N \)  is essential to observe the anti-PT phase transition. As in this way, we can vary  \(  \kappa  \)  by changing  \( N \)  but keeping  \(  \Delta k \)  unchanged. At the end of the MOT loading time in each cycle, all the atoms are optically pumped into $|1 \rangle$. In the following FWM duty time window of each cycle, the co-propagating pump (left circular polarization \(  \sigma ^{-} \), 780 nm) and coupling ( \(  \sigma ^{+} \), 795 nm) lasers are collimated with the same beam diameter of  \( 1.54 \)  mm and aligned at an intersection angle of  \(  \theta =0.2^{ ^{\circ} } \)  to the longitudinal  \( z \) axis [Fig.~\ref{fig:figure1}(b)]. Introducing this small  \(  \theta  \)  is to set an optimally fixed phase-mismatch value, \(  \Delta k=97.2 \)  rad/m, allowed in our experiment. The seeded Stokes light (\(  \sigma ^{-} \)) is then applied along the  \( +z \) -axis with a waist diameter of 300 $ \mu $m at the MOT center to create its twin beam, anti-Stokes (\(  \sigma ^{+} \)), and vice versa. In practice, Stokes experiences inevitably small linear Raman amplification due to the applied pump laser, while anti-Stokes always suffers the unavoidable linear absorption owing to imperfect EIT caused by nonzero  \(  \gamma _{12} \) . Though it is impossible to completely remove this loss and gain in reality, through optimization we have substantially reduced both quantities more than 10 times with respect to \(  \Delta k \) (Fig.~\ref{fig:fig2}).

The dynamics of anti-PT eigenmodes can be experimentally probed via measuring the spectral transmission profiles ($|A\left(\Delta\omega\right)|^{2}$ to $|D\left(\Delta  \omega\right)|^{2}$) of the Stokes and anti-Stokes channels by changing  \( N \). In this work, we maintain the pump and coupling Rabi frequencies at  \(  \Omega _{p}=2 \pi  \times 5.0 \)  MHz and  \(  \Omega _{c}=2 \pi  \times 8.7 \)  MHz, respectively.  \(  \vert A \left(  \Delta  \omega  \right)  \vert ^{2} \)  and  \(  \vert C \left(  \Delta  \omega  \right)  \vert ^{2} \)  are obtained via measuring the normalized anti-Stokes and Stokes output powers with respect to the anti-Stokes seeding power; whereas  \(  \vert B \left(  \Delta  \omega  \right)  \vert ^{2} \)  and  \(  \vert D \left(  \Delta  \omega  \right)  \vert ^{2} \)  are attained by normalizing their outputs to the seeding Stokes power. In Figs.~\ref{fig:fig3}(a) and (b), two sets of representative snapshots of spectral transmissions are displayed as the function of  \(  \Delta  \omega  \)  for different  \(  \delta  \)  and  \( N \). The normalized outputs at  \(  \Delta  \omega =0 \)  thus render the information of  \(  \lambda _{ \pm } \)  and the evolution of the corresponding eigenmodes. Away from the two-photon resonance (\(  \Delta  \omega  \neq 0 \)), the linear EIT loss and Raman gain becomes important and deviates the nonlinear interaction away from anti-PT due to the propagation effect. By taking into account this effect, as one can see, the experimental data shows a good agreement with the theoretical curves (solid lines) in the whole spectrum.

As anti-PT symmetry is more evident at  \(  \Delta  \omega =0 \), we extract the corresponding transmission strengths from the measured spectra at that location in terms of  \( N \)  and  \(  \delta  \). For instance, Figs.~\ref{fig:fig4}(a1) and (b1) illustrate, respectively, the variations of the normalized Stokes and anti-Stokes outputs subject to different seeding; while Figs. 4(a2) and (b2) depict their individual ratios with respect to the total output power. Theoretically, below the degeneracy (\(  \delta <1 \)), in addition to the bounded power, a power oscillatory behavior is also anticipated to arise between two fields for a sufficient medium length. Owing to the restricted MOT length (\( L=1.5 \) cm) and relatively large  \(  \Delta k \), unfortunately, only a minute fraction of power oscillation can be recognized in the current architecture. For comparison, the numerical results for a medium with a larger length (10\( L=15 \) cm) are presented in Figs. 4(a2) and (b2) by the dashed lines, which not only indicate the occurrence of the phase transition at  \(  \delta =1 \)  or  \( N=7.90 \times 10^{16} \)  m\textsuperscript{-3} but also showcase clearly the oscillations. In contrary, after the threshold both output intensities shall augment exponentially even for a finite  \( L \)  as the FFWM interaction is enforced by anti-PT symmetry. Indeed, the experiment confirms such amplification as indicated in Figs.~\ref{fig:fig4}(a1) and (b1). Additionally, one would expect the intensity ratio between the two beams to approach unit after certain propagation distance. This is to some extent affirmed in Figs.~\ref{fig:fig4}(a2) and (b2), where the slow convergence is partly due to the limited medium length and the unavoidably small EIT loss (and Raman gain). Interestingly, in this phase-unbroken regime, anti-PT symmetry automatically compensates large  \(  \Delta k \)  and hence provides a new way for efficient nonlinear conversion by the relaxation of desired phase matching. In the experiment, we also carefully adjust the seeding powers to be only 1 nW (\(  \Omega _{s}=2 \pi  \times 0.053 \)  MHz and  \(  \Omega _{as}=2 \pi  \times 0.065 \)  MHz).

From the output transmission spectra (Fig.~\ref{fig:fig3}), one can obtain the phase constants (Re$\left[  \lambda _{ \pm } \right]$) as well as attenuation/amplification constants (Im$\left[  \lambda _{ \pm } \right]$) as functions of  \( N \). Since the system parameters make the anti-PT phase transition more unequivocal at \(  \Delta  \omega =0 \) , the displayed  Re$\left[  \lambda _{ \pm } \right]$ and Im$\left[  \lambda _{ \pm } \right]$ in Fig.~\ref{fig:fig5} are attained at this two-photon resonance. In spite of the system’s imperfections, the phase-transition threshold (\( N=7.90 \times 10^{16} \)  m$^{-3}$) is still accessible in our experiment for seeding either anti-Stokes Fig.~\ref{fig:fig5}(a) or Stokes (b). According to Eq. (\ref{eq:eigenvalueEq2}), below the transition the eigen-propagation constants shall be dominated by the phase constants while above the EP, the attenuation and amplification constants start to play the major role. This is also well confirmed from the fitting theoretical curves to the experimental data. We have further looked at anti-PT behaviors beyond  \(  \Delta  \omega =0 \). As one can see, the essential features of anti-PT symmetry is still noticeable even though the EIT loss becomes comparable to  \(  \Delta k \).

In summary, the demonstrated anti-PT symmetric nonlinear optics is fundamentally different from previous results. Of importance, as no appreciable linear gain and loss involved, such a scheme can be readily extended to the quantum optics domain without Langevin noises \cite{47,48, 50} that is challenging to standard gain-loss PT/anti-PT systems \cite{2,3,4,5,6,15}. All these attributes are believed to deserve further investigations and explorations on counter-intuitive optical phenomena as well as a new generation of anti-PT-enabled optical devices for quantum information application.

\begin{acknowledgments}
The work was supported by the Hong Kong Research Grants Council (Project Nos. 16308118 and C6005-17G) and in part by the matching funding by The Hong Kong University of Science and Technology (Project No. VPRGO18SC08). Y.-H. Z. and J.W. were supported by the US NSF EFMA-1741693, US NSF 1806519, and Kennesaw State University.
\end{acknowledgments}

\bibliography{anti-PT}

\end{document}